\documentclass[12pt]{article}
\usepackage{enumerate}
\usepackage{amsfonts}
\usepackage{amsmath}
\usepackage{amssymb}

\newcounter{defin}  \newcounter{lemma}  \newcounter{theorem}
\newcounter{property} \newcounter{corol}  \newcounter{remark}

\newenvironment{lemma}{\par\refstepcounter{lemma}
     \textbf{Lemma \thelemma.} }{\rm\par}
\newenvironment{theorem}{\par\refstepcounter{theorem}
     \textbf{Theorem \thetheorem.}\ }{\rm\par}
\newenvironment{property}{\par\refstepcounter{property}
     \textbf{Proposition \theproperty.}\ }{\rm\par}
\newenvironment{corollary}{\par\refstepcounter{corol}
     \textbf{Corollary \thecorol.} }{\rm\par}

\newenvironment{remark}{\par\refstepcounter{remark}
     \textbf{Remark \theremark.}}{\rm\par}

\begin{document}

\title{Properties of probability measures on the set of quantum states and their applications.}
\author{M.E.Shirokov \thanks{Steklov Mathematical Institute, 119991 Moscow,
Russia, e-mail:msh@mi.ras.ru}}
\date{}
\maketitle

\begin{abstract}
Two basic properties of the set of all probability measures on the
set of quantum states and their corollaries  are considered. Several
applications of these properties to analysis of functional
constructions widely used in quantum information theory are
discussed.
\end{abstract}

\section{Introduction}

The notion of an ensemble as a collection of quantum states with
corresponding probability distribution is widely used in quantum
information theory. In particular, several important characteristics
such as the Holevo capacity of a quantum channel and the
Entanglement of Formation of a  state of a composite quantum system are
defined by optimization of the particular functionals depending on
ensemble of quantum states \cite{B&Ko,H-SSQT, N&Ch}.

An ensemble of quantum states can be considered as an atomic
probability measure on the set of all quantum states, whose atoms
correspond to the states of the ensemble. So, it is natural to
consider an arbitrary Borel probability measure on the set of all
quantum  states as a generalized ensemble. This point of view is
especially useful in dealing with infinite dimensional quantum
channels and systems, since in this case it is necessary to consider
continuous ensembles of states, t.i. families of states indexed by a
real valued parameter \cite{H-Sh-2}. One of the advantages of this
approach is based on possibility to use general results of the theory of
probability measures on complete separable metric spaces
\cite{Bil,Par}.

In this paper we focus attention on the following two basic
properties of the set of all probability measures on the set of
quantum states endowed with the weak convergence topology:
\begin{itemize}
    \item compactness of a subset of probability measures,
    whose barycenters form compact subset of states;
    \item openness of the barycenter map and of its restrictions to the subsets of measures supported by states with bounded maximal rank.
\end{itemize}
In fact, these properties, described in detail in Section 3 and 4,
reflect the special relations between the topology and convex
structure of the set of quantum states. The first of them can be
considered as a kind of weak compactness since it provides
generalization to the set of quantum states of several results well
known for compact convex sets. The second one shows roughly speaking
that any small perturbation of the average state of an ensemble of
quantum states can be realized by appropriate small perturbations of
the states of this ensemble, it gives possibility, in particular, to
prove preserving of upper semicontinuity of a function under taking the
convex closure and the convex roof (typical operations of convex
analysis).\medskip

Several applications of the above properties are considered in
Section 5.

\section{Preliminaries}

Let $\mathcal{H}$ be a separable Hilbert space,
$\mathfrak{B}(\mathcal{H})$ -- the set of all bounded operators in
$\mathcal{H}$ with the cone $\mathfrak{B}_{+}(\mathcal{H})$ of all
positive operators, $\mathfrak{T}( \mathcal{H})$ -- the Banach space
of all trace-class operators with the trace norm
$\Vert\cdot\Vert_{1}$ and $\mathfrak{S}(\mathcal{H})$ -- the closed
convex subset of $\mathfrak{T}(\mathcal{H})$ consisting of all
positive operators with the unit trace -- density operators in
$\mathcal{H}$, which is complete separable metric space with the
metric defined by the trace norm. Each density operator uniquely
defines a normal state on $\mathfrak{B}(\mathcal{H})$ \cite{B&R},
so, in what follows we will also for brevity use the term "state".

Let $\mathfrak{S}_k(\mathcal{H})$ be the subset of $\mathfrak{S}(\mathcal{H})$ consisting of states with rank $\leq k\in\mathbb{N}$.
In particular, $\mathfrak{S}_1(\mathcal{H})$ is the set of all pure states -- one-dimensional projectors.

We denote by $\mathrm{co}(\mathcal{A})$ and $\overline{\mathrm{co}}(\mathcal{A})$ the convex hull and the convex closure of a set
$\mathcal{A}$.  The set of all extreme points of a convex set
$\mathcal{A}$ will be denoted $\mathrm{extr}\mathcal{A}$.

Let $C(\mathcal{A})$ be the set of all continuous bounded functions on a set $\mathcal{A}$.

For an arbitrary function $f$ on the set $\mathfrak{S}(\mathcal{H})$
we denote by $\,\mathrm{co}f\,$ and $\,\overline{\mathrm{co}}f\,$
its convex hull (defined as the maximal convex function majorized by
$f$) and its convex closure (defined as the maximal convex lower
semicontinuous function majorized by $f$) correspondingly
\cite{J&T}.

For an arbitrary closed subset $\mathcal{A}$ of
$\mathfrak{S}(\mathcal{H})$ let $\mathcal{P}(\mathcal{A})$ be the
set of all Borel probability measures on $\mathcal{A}$ endowed with
the topology of weak convergence \cite{Bil,Par}. Since $\mathcal{A}$
is a complete separable metric space, $\mathcal{P}(\mathcal{A})$ can
be considered as a complete separable metric space as well
\cite{Par}. Let $\mathcal{P}^{\,a}(\mathcal{A})$ and
$\mathcal{P}^{f}(\mathcal{A})$ be the subsets of
$\mathcal{P}(\mathcal{A})$ consisting of atomic measures with
countable and finite number of atoms correspondingly.

An atomic measure in $\mathcal{P}(\mathcal{A})$ consisting of the atoms
$\{\rho_{i}\}\subset\mathcal{A}$ with the corresponding weights
$\{\pi_{i}\}$ will be denoted $\{\pi_{i},\rho_{i}\}$. From the
physical point of view it can be interpreted as a discrete ensemble of quantum states while an arbitrary measure in
$\mathcal{P}(\mathcal{A})$ -- as a continuous ensemble.

The \textit{barycenter} of a measure
$\mu\in\mathcal{P}(\mathcal{A})$ is the state in
$\overline{\mathrm{co}}(\mathcal{A})$ defined by the Bochner
integral
\[
\textbf{b}(\mu)=\int_{\mathcal{A}}\sigma \mu(d\sigma).
\]
If $\mu=\{\pi_{i},\rho_{i}\}$ then
$\textbf{b}(\mu)=\sum_i\pi_{i}\rho_{i}$ is the average state of the
ensemble  $\{\pi_{i},\rho_{i}\}$.\smallskip

For an arbitrary subset $\mathcal{B}$ of $\,\overline{\mathrm{co}}
(\mathcal{A})$ we denote by
$\mathcal{P}_{\mathcal{B}}(\mathcal{A})$ the subset of
$\mathcal{P}(\mathcal{A})$ consisting of measures with the
barycenter in $\mathcal{B}$.


\section{Compactness criterion and its corollaries}

The set $\mathcal{P}(\mathfrak{S}(\mathcal{H}))$ is compact if and
only if $\dim\mathcal{H}<+\infty$ (since the set
$\mathfrak{S}(\mathcal{H})$ is compact if and only if
$\dim\mathcal{H}<+\infty$). But in the case
$\dim\mathcal{H}=+\infty$ the following compactness criterion for
subsets of $\mathcal{P}(\mathfrak{S}(\mathcal{H}))$ can be proved by
using Prokhorov's theorem \cite[Proposition 2]{H-Sh-2}.\vspace{5pt}

\begin{theorem}\label{th-1}
\textit{The set
$\,\mathcal{P}_{\mathcal{A}}(\mathfrak{S}(\mathcal{H}))\subset\mathcal{P}(\mathfrak{S}(\mathcal{H}))$
is compact if and only if the set
$\mathcal{A}\subset\mathfrak{S}(\mathcal{H})$ is compact.}
\end{theorem}\vspace{5pt}

The property stated in Theorem \ref{th-1} is not purely topological
but it reflects the particular relation between the topology and the
convex structure of the set $\mathfrak{S}(\mathcal{H})$ (the convex
structure is involved in the definition of the barycenter of a
probability measure). This property is studied in \cite{P&Sh} in the
context of arbitrary closed complete metrizable bounded subsets of a
separable locally convex topological space, where it is called the
$\mu$-compactness property. It turns out that $\mu$-compact convex
sets (in particular, the set $\mathfrak{S}(\mathcal{H})$) inherit
several well known  properties of compact convex sets such as the
Choquet theorem of barycentric representation, lower semicontinuity
of the convex hull of any continuous bounded function, etc.

Note first the following two versions of the Choquet decomposition
for closed convex subsets of
$\mathfrak{S}(\mathcal{H})$ (see \cite[Proposition 5]{P&Sh} and
\cite[Lemma 1]{Sh-11}).\vspace{5pt}

\begin{corollary}\label{c-1}
A) \textit{Let $\mathcal{A}$ be a closed convex subset of
$\,\mathfrak{S}(\mathcal{H})$. Then every state in $\mathcal{A}$ can
be represented as the barycenter of some Borel probability measure
supported by the closure of $\,\mathrm{extr}\mathcal{A}$.}\medskip

B) \textit{Let $\mathcal{A}$ be a closed subset of
$\,\mathfrak{S}(\mathcal{H})$. Then every state in
$\,\overline{\mathrm{co}}(\mathcal{A})$ can be represented as the
barycenter of some Borel probability measure supported by
the set $\mathcal{A}$.}
\end{corollary}\vspace{5pt}

Corollary \ref{c-1}B shows, in particular, that
$\,\mathrm{extr}(\overline{\mathrm{co}}(\mathcal{A}))\subseteq\mathcal{A}\,$
for an arbitrary closed subset
$\mathcal{A}\subseteq\mathfrak{S}(\mathcal{H})$, t.i. \emph{no
extreme points can appear by taking closure of a convex hull of a
closed subset of quantum states}.\footnote{This does not hold in
general for closed subsets of a noncompact set. Indeed, let
$\mathcal{A}=\{e_i\}_{i=1}^{+\infty}$ be an orthonormal basis in a
Hilbert space, then it is easy to see that
$0\in\mathrm{extr}(\overline{\mathrm{co}}(\mathcal{A}))$.}\vspace{5pt}

For an arbitrary state $\rho$ in $\mathfrak{S}(\mathcal{H})$ the set
$\mathcal{P}^f_{\{\rho\}}(\mathfrak{S}(\mathcal{H}))$ is a dense
subset of $\mathcal{P}_{\{\rho\}}(\mathfrak{S}(\mathcal{H}))$
\cite[Lemma 1]{H-Sh-2}. By using Theorem \ref{th-1} this simple
result can be strengthened as follows \cite[Lemma
5]{Sh-11}.\vspace{5pt}

\begin{corollary}\label{c-2} \textit{For an arbitrary state $\,\rho\in\mathfrak{S}(\mathcal{H})\,$
and natural $\,k\,$ the set
$\mathcal{P}_{\{\rho\}}^{\,a}(\mathfrak{S}_{k}(\mathcal{H}))$ is a
dense subset of
$\,\mathcal{P}_{\{\rho\}}(\mathfrak{S}_{k}(\mathcal{H}))$. }
\end{corollary}
\vspace{5pt}

This means that any probability measure supported by the set of
states of rank $\leq k$ can be weakly approximated by some sequence
of atomic measures -- countable ensembles of states of rank $\leq k$
with the same barycenter.\vspace{5pt}

An important implication of the compactness criterion in Theorem
\ref{th-1} is contained in the following assertion \cite[Lemma
2A]{Sh-11}.\vspace{5pt}

\begin{corollary}\label{c-3}
\textit{Let $f$ be a lower semicontinuous lower bounded function on
a closed subset $\mathcal{A}$ of $\,\mathfrak{S}(\mathcal{H})$.}
\textit{Then the convex function
$$
\check{f}_{\mathcal{A}}(\rho)\doteq\inf_{\mu\in
\mathcal{P}_{\{\rho\}}(\mathcal{A})}\int_{\mathcal{A}}
f(\sigma)\mu(d\sigma)
$$
is lower semicontinuous on the set
$\,\overline{\mathrm{co}}(\mathcal{A})$.\footnote{Correctness of the
definition of this function follows from Corollary \ref{c-1}.} For
an arbitrary state $\rho$ in $\,\overline{\mathrm{co}}(\mathcal{A})$
the infimum in the definition of the value
$\check{f}_{\mathcal{A}}(\rho)$ is achieved at a particular measure
in $\mathcal{P}_{\{\rho\}}(\mathcal{A})$.}\vspace{5pt}
\end{corollary}\vspace{5pt}

It is well known that for an arbitrary increasing sequence
$\{f_{n}\}$ of continuous functions on a convex compact set
$\mathcal{A}$ the corresponding sequence
$\{\overline{\mathrm{co}}f_{n}\}$ pointwise converges to the
function $\overline{\mathrm{co}}f_{0}$, where
$f_{0}=\sup_{n}f_{n}$.\footnote{If $f_{0}$ is a continuous function
then this follows from Dini's lemma. The importance of the
compactness condition can be shown by the sequence of the functions
$f_{n}(x)=\exp({-x^{2}/n})$ on $\mathbb{R}$, converging to the
function $f_{0}(x)\equiv 1$, such that
$\overline{\mathrm{co}}f_{n}(x)\equiv 0$ for all $n$.} It turns out
that the compactness criterion in Theorem \ref{th-1} implies (in
fact, \textit{means}, see Remark \ref{contrversion} below) the
analogous observation for the noncompact set
$\mathcal{A}=\mathfrak{S}(\mathcal{H})$ \cite[Proposition
6]{Sh-9}.\vspace{5pt}

\begin{corollary}\label{c-3+}
\textit{For an arbitrary increasing sequence $\{f_{n}\}$ of lower
semicontinuous lower bounded functions on the set
$\,\mathfrak{S}(\mathcal{H})$ and an arbitrary converging sequence
$\,\{\rho_{n}\}$ of states in $\,\mathfrak{S}(\mathcal{H})$ the
following inequality holds
$$
\liminf_{n\rightarrow+\infty}\overline{\mathrm{co}}f_{n}(\rho_{n})\geq\overline{\mathrm{co}}f_{0}(\rho_{0}),\quad
\textit{where}\quad f_{0}=\sup_{n}f_{n}\quad\textit{and}\quad
\rho_{0}=\lim_{n\rightarrow+\infty}\rho_{n}.
$$}
\textit{In particular,
$$
\lim_{n\rightarrow+\infty}\overline{\mathrm{co}}f_{n}(\rho)=\overline{\mathrm{co}}f_{0}(\rho),\quad
\forall \rho\in\mathfrak{S}(\mathcal{H}).
$$}
\end{corollary}\vspace{5pt}

\begin{remark}\label{contrversion}
The compactness criterion in Theorem \ref{th-1} can be derived from
validity of the last assertion of Corollary \ref{c-3+} for any
increasing  sequence $\{f_{n}\}\subset C(\mathfrak{S}(\mathcal{H}))$
converging to a function $f_{0}\in C(\mathfrak{S}(\mathcal{H}))$
\cite{Sh-9}.
\end{remark}

\section{Openness of the barycenter map}

The barycenter map
\begin{equation}\label{b-map}
 \mathcal{P}(\mathfrak{S}(\mathcal{H}))\ni\mu\mapsto\textbf{b}(\mu)\in \mathfrak{S}(\mathcal{H})
\end{equation}
is a continuous surjection \cite{H-Sh-2}. An important property of
this map is presented in the following theorem (see Section 3 in
\cite{Sh-11}).\vspace{5pt}

\begin{theorem}\label{th-2}
A) \textit{The barycenter map (\ref{b-map}) and its restriction to
the set $\mathcal{P}^{\,a}(\mathfrak{S}(\mathcal{H}))$ are open
surjections.}\medskip

B) \textit{The restrictions of the barycenter map (\ref{b-map}) to
the sets $\,\mathcal{P}(\mathfrak{S}_k(\mathcal{H}))$ and
$\,\mathcal{P}^{\,a}(\mathfrak{S}_k(\mathcal{H}))$ are open
surjections for each $\,k\in\mathbb{N}$.}
\end{theorem}\vspace{5pt}

Physically openness of the map
$\mathcal{P}(\mathfrak{S}(\mathcal{H}))\ni\mu\mapsto\textbf{b}(\mu)\in\mathfrak{S}(\mathcal{H})$
(correspondingly, of the map
$\mathcal{P}^{\,a}(\mathfrak{S}(\mathcal{H}))\ni\mu\mapsto\textbf{b}(\mu)\in\mathfrak{S}(\mathcal{H})$)
means, roughly speaking, that any small perturbation of the average
state of a given continuous (correspondingly, of discrete) ensemble
of states  can be realized by appropriate small perturbations of the
states of this ensemble. By the
$\mu$\nobreakdash-\hspace{0pt}compact generalization of the
Vesterstrom-O'Brien theorem (\cite[Theorem 1]{P&Sh}) this property
is equivalent to the following ones:
\begin{itemize}
    \item the map $\;\mathfrak{S}(\mathcal{H})\times\mathfrak{S}(\mathcal{H})\ni (\rho,\sigma)\ \mapsto\
    \frac{1}{2}(\rho + \sigma)\in \mathfrak{S}(\mathcal{H})\;$ is open;
    \item the map
    $\;\mathcal{P}(\mathfrak{S}_1(\mathcal{H}))\ni\mu\mapsto\textbf{b}(\mu)\in\mathfrak{S}(\mathcal{H})\;$ is
    open;
    \item $\mathrm{co}f=\overline{\mathrm{co}}f\in C(\mathfrak{S}(\mathcal{H}))\;$ for any concave function $\,f\,\in
    C(\mathfrak{S}(\mathcal{H}))$;
    \item $\mathrm{co}f=\overline{\mathrm{co}}f\in C(\mathfrak{S}(\mathcal{H}))\;$ for any function $\,f\,\in
    C(\mathfrak{S}(\mathcal{H}))$.
 \end{itemize}

According to the terminology accepted in the convex analysis (cf.
\cite{Grzaslewicz,Stefania}) the first of the above properties can
be called \emph{stability} of the set $\mathfrak{S}(\mathcal{H})$.
It is essential that the first assertion of Theorem \ref{th-2} can
be deduce from stability of the set $\mathfrak{S}(\mathcal{H})$
without using $\mu$-compactness of this set.

The second assertion of Theorem \ref{th-2} shows that any small
perturbation of the average state of a given (discrete or
continuous) ensemble can be realized by appropriate small
perturbations of the states of this ensemble \textit{without
increasing of the maximal rank of these states}. In \cite{Sh-11} it
is called  the \textit{strong stability property} of the set
$\mathfrak{S}(\mathcal{H})$.\vspace{5pt}

Theorem \ref{th-2} and Lemma 2B in \cite{Sh-11} imply the following
observation.
\medskip
\begin{corollary}\label{c-5}
\textit{Let $\mathcal{A}$ be either $\,\mathfrak{S}(\mathcal{H})$ or
$\,\mathfrak{S}_k(\mathcal{H})$ for $\,k\in\mathbb{N}$. Let $f$ be a
lower semicontinuous lower bounded function on the set
$\mathcal{A}$.} \textit{Then the concave function
$$
\hat{f}_{\mathcal{A}}(\rho)\doteq\sup_{\mu\in
\mathcal{P}_{\{\rho\}}(\mathcal{A})}\int_{\mathcal{A}}
f(\sigma)\mu(d\sigma)
$$
is lower semicontinuous on the set $\,\mathfrak{S}(\mathcal{H})$.
The supremum in the definition of the value
$\hat{f}_{\mathcal{A}}(\rho)$ can be taken over the set
$\,\mathcal{P}^f_{\{\rho\}}(\mathcal{A})$ if
$\,\mathcal{A}=\mathfrak{S}(\mathcal{H})$ and over the set
$\,\mathcal{P}^{\,a}_{\{\rho\}}(\mathcal{A})$ if
$\,\mathcal{A}=\mathfrak{S}_k(\mathcal{H})$.}\vspace{5pt}
\end{corollary}\vspace{5pt}

The last assertion of this corollary follows form lower
semicontinuity of the functional $\mu\mapsto\int_{\mathcal{A}}
f(\sigma)\mu(d\sigma)$ used with Lemma 1 in \cite{H-Sh-2} if
$\mathcal{A}=\mathfrak{S}(\mathcal{H})$ and with Corollary \ref{c-2}
if $\mathcal{A}=\mathfrak{S}_k(\mathcal{H})$. It shows that
$$
\hat{f}_{\mathcal{A}}(\rho)=\sup_{\{\pi_i,\rho_i\}\in
\mathcal{P}^{\,*}_{\{\rho\}}(\mathcal{A})}\sum_i\pi_i f(\rho_i),
$$
where $\,*=f\,$ if $\,\mathcal{A}=\mathfrak{S}(\mathcal{H})\,$ and
$\,*=a\,$ if $\,\mathcal{A}=\mathfrak{S}_k(\mathcal{H})$.

\section{Applications}

\subsection{Approximation of concave and convex lower semicontinuous functions}

It is well known that any lower semicontinuous lower bounded
function on a metric space $\mathcal{A}$ can be represented as a
pointwise limit of incresing sequence of continuous bounded
functions \cite{Ali}. The $\mu$-compactness and stability of the set
$\mathfrak{S}(\mathcal{H})$ (expressed in Theorems \ref{th-1} and
\ref{th-2}A respectively) imply the following assertion.
\vspace{5pt}

\begin{property}\label{approxim}
\emph{Let $f$ be a concave (corresp. convex) lower semicontinuous
lower bounded function on $\,\mathfrak{S}(\mathcal{H})$. Then
$f=\sup_n f_n$, where $\{f_n\}$ is a nondecreasing sequence of
concave (corresp. convex) functions in
$C(\mathfrak{S}(\mathcal{H}))$.}
\end{property}\vspace{5pt}

\textbf{Proof.} Let $f$ be a concave function and $\{g_n\}$  be a
nondecreasing sequence in $C(\mathfrak{S}(\mathcal{H}))$ such that
$f=\sup_n g_n$. Let $f_n$ be a concave hull of $\,g_n$ (that is
$f_n=-\mathrm{co}(-g_n)$). By
$\mu$\nobreakdash-\hspace{0pt}compactness and stability of the set
$\mathfrak{S}(\mathcal{H})$ and the generalized Vesterstrom-O'Brien
theorem (Theorem 1 in \cite{P&Sh}) the sequence $\{f_n\}$ belongs to
the set $C(\mathfrak{S}(\mathcal{H}))$. Since $g_n\leq f_n$, we have
$f=\sup_n f_n$.

Let $f$ be a convex function and $\{g_n\}$  be a nondecreasing
sequence in $C(\mathfrak{S}(\mathcal{H}))$ such that $f=\sup_n g_n$.
Let $f_n$ be a convex hull of $\,g_n$. By
$\mu$\nobreakdash-\hspace{0pt}compactness and stability of the set
$\mathfrak{S}(\mathcal{H})$ and the generalized Vesterstrom-O'Brien
theorem (Theorem 1 in \cite{P&Sh}) the sequence $\{f_n\}$ belongs to
the set $C(\mathfrak{S}(\mathcal{H}))$. By Corollary \ref{c-3+} we
have $f=\sup_n f_n$. $\square$

\subsection{Several results concerning the Choquet ordering}

Consider the following partial ordering on the set
$\mathcal{P}(\mathfrak{S}(\mathcal{H}))$. We say that $\mu\succ\nu$
if and only if
\begin{equation}\label{order-ineq}
\int_{\mathfrak{S}(\mathcal{H})}f(\sigma)\mu(d\sigma)
\geq\int_{\mathfrak{S}(\mathcal{H})}f(\sigma)\nu(d\sigma)
\end{equation}
for any convex function $f$ in $C(\mathfrak{S}(\mathcal{H}))$.

This partial ordering is often called the Choquet ordering
\cite{Phelps}. It coincides with some others partial orderings on
the set $\mathcal{P}(\mathfrak{S}(\mathcal{H}))$, in particular,
with the delation ordering \cite{Bourgin,Edgar}.\smallskip

Note first the following simple corollary of Proposition
\ref{approxim}.\vspace{5pt}

\begin{lemma}\label{order-ineq+}
\textit{Let $\mu$ and $\nu$ be measures in
$\mathcal{P}(\mathfrak{S}(\mathcal{H}))$ such that $\,\mu\succ\nu$.
Then inequality (\ref{order-ineq}) holds for any convex function $f$
on the set $\,\mathfrak{S}(\mathcal{H})$, which is either lower
semicontinuous or upper semicontinuous and upper bounded.}
\end{lemma}\vspace{5pt}

To derive this assertion from Proposition \ref{approxim} it suffices
to note that any convex lower semicontinuous function on the set
$\mathfrak{S}(\mathcal{H})$ is either lower bounded or does not take
finite values (see Lemma 2 in \cite{Sh-9}) and to use the monotone
convergence theorem.

It is easy to see (by considering affine continuous functions on
$\mathfrak{S}(\mathcal{H})$) that the relation $\,\mu\succ\nu\,$
implies $\,\textbf{b}(\mu)=\textbf{b}(\nu)$.

Intuitively speaking, the relation $\mu\succ\nu$ means that "the
mass of $\mu$ is removed farther away from the common barycenter of
$\mu$ and $\nu$, and comes close to the extreme boundary"
\cite{Alf}. Note that the extreme boundary (the set of extreme
points) of the set $\mathfrak{S}(\mathcal{H})$ coincides with  the
set of all pure states (states of rank 1) and that for an arbitrary
subspace $\mathcal{H}_{0}$ of the space $\mathcal{H}$ the subset
$\mathfrak{S}(\mathcal{H}_{0})$ is a \textit{face} of the set
$\mathfrak{S}(\mathcal{H})$ \cite{J&T}. Thus the above
characterization of the partial ordering $"\succ"$ is confirmed by
the following observations.
\medskip

\begin{property}\label{ppm-p-1}
\textit{Let $\mu$ and $\nu$ be arbitrary measures in
$\mathcal{P}(\mathfrak{S}(\mathcal{H}))$ such that $\mu\succ\nu$.
Then}
\begin{itemize}
  \item \textit{$\mu(\mathcal{A})\geq\nu(\mathcal{A})$ for any Borel subset $\mathcal{A}\subseteq\mathfrak{S}_1(\mathcal{H})$;}
  \item \textit{$\mu(\mathfrak{S}(\mathcal{H}_{0}))\geq\nu(\mathfrak{S}(\mathcal{H}_{0}))$
  for any subspace $\mathcal{H}_{0}\subseteq\mathcal{H}$;}
  \item \textit{$\mu(\mathfrak{S}_{k}(\mathcal{H}))\geq\nu(\mathfrak{S}_{k}(\mathcal{H}))$ for any $\,k\in\mathbb{N}$.}
\end{itemize}
\end{property}\medskip

\textbf{Proof.} The first assertion of the proposition for any
closed subset $\mathcal{A}$ follows from Lemma \ref{order-ineq+},
since the indicator function of this set is obviously convex and
upper semicontinuous. To complete its proof it suffices to note that
every measure in $\mathcal{P}(\mathfrak{S}(\mathcal{H}))$ is a Radon
measure.

The second and the third assertions also  follow from Lemma
\ref{order-ineq+}, since the indicator functions of the sets
$\mathfrak{S}(\mathcal{H}_{0})$ and $\mathfrak{S}_{k}(\mathcal{H})$
are also convex and upper semicontinuous. $\square$\medskip

According to \cite{Phelps} a measure $\mu$ in
$\mathcal{P}(\mathfrak{S}(\mathcal{H}))$ is called \textit{maximal}
if $\,\nu\succ\mu\,$ implies $\,\nu=\mu\,$ for any measure $\nu$ in
$\mathcal{P}(\mathfrak{S}(\mathcal{H}))$.\medskip

\begin{corollary}\label{ppm-c-n}
\textit{The set of all maximal measures in
$\mathcal{P}(\mathfrak{S}(\mathcal{H}))$ coincides with
$\mathcal{P}(\mathfrak{S}_1(\mathcal{H}))$.}

\textit{For every measure $\mu$ in
$\mathcal{P}(\mathfrak{S}(\mathcal{H}))$ there exists a measure
$\hat{\mu}$ in $\mathcal{P}(\mathfrak{S}_1(\mathcal{H}))$ such that
$\,\hat{\mu}\succ\mu$.}
\end{corollary}\medskip

\textbf{Proof.} The assertions of this corollary can be deduced from
general results of the theory of measures on convex sets
\cite{Bourgin,Edgar}, but we want to show that the above-stated
properties of the set $\mathcal{P}(\mathfrak{S}(\mathcal{H}))$
provide a very simple and \emph{constructive} way of their proof.

Let $\mu$ be an arbitrary measure in
$\mathcal{P}(\mathfrak{S}_1(\mathcal{H}))$. By Proposition
\ref{ppm-p-1} the assumption $\nu\succ\mu$ for some measure $\nu$ in
$\mathcal{P}(\mathfrak{S}(\mathcal{H}))$ implies
$\nu(\mathcal{A})\geq\mu(\mathcal{A})$ for any Borel set
$\mathcal{A}$ of pure states. Since $\mu$ and $\nu$ are
\textit{probability} measures and $\mu$ is supported by pure states,
equality necessarily holds in the above inequality, and hence
$\mu=\nu$. Thus $\mu$ is a maximal measure in
$\mathcal{P}(\mathfrak{S}(\mathcal{H}))$.

If $\mu$ is a maximal measure in
$\mathcal{P}(\mathfrak{S}(\mathcal{H}))$ then, by the below
observation, there exists a measure $\hat{\mu}$ in
$\mathcal{P}(\mathfrak{S}_1(\mathcal{H}))$ such that
$\hat{\mu}\succ\mu$ and hence $\mu=\hat{\mu}$.

Let $\mu$ be an arbitrary measure in
$\mathcal{P}(\mathfrak{S}(\mathcal{H}))$.  By Lemma 1 in
\cite{H-Sh-2} there exists a sequence $\{\mu_{n}\}$ of measures in
$\mathcal{P}^{f}_{\{\textbf{b}(\mu)\}}(\mathfrak{S}(\mathcal{H}))$
converging to the measure $\mu$. Decomposing each atom of the
measure $\mu_{n}$ into a convex combination of pure states we obtain
(as in the proof of the Theorem in \cite{H-Sh-2}) the measure
$\hat{\mu}_{n}$ with the same barycenter supported by pure states.
It is easy to see that $\hat{\mu}_{n}\succ\mu_{n}$. By Theorem
\ref{th-1} the sequence $\{\hat{\mu}_{n}\}_{n>0}$ is  relatively
compact and hence it contains a subsequence $\{\hat{\mu}_{n_{k}}\}$
converging to a particular measure $\hat{\mu}$ supported by pure
states. Since $\hat{\mu}_{n_{k}}\succ\mu_{n_{k}}$ for all $k$, the
definition of the weak convergence implies $\hat{\mu}\succ\mu$.
$\square$

\subsection{On convex (concave) envelopes of a function}

The convex closure\footnote{the lower (convex) envelope in terms of
\cite{Alf}.} $\overline{\mathrm{co}}f$ of a function $f$ on a convex
topological space $X$ is defined as the maximal closed (lower
semicontinuous) convex function on $X$ majorized by $f$ and
generally does not coincide with the convex hull $\mathrm{co}f$ of
the function $f$ defined as the maximal convex function on $X$
majorized by $f$ \cite{Alf,J&T}. These notions (in the case
$X=\mathfrak{S}(\mathcal{H})$) play essential roles in  quantum
information theory, since they are involved (sometimes implicitly)
in definitions of several important characteristics of quantum
systems and channels \cite{A&B,B&Ko,H-Sh-2}.\smallskip

In this subsection we present several results concerning the notions
of a convex closure and of a convex hull of a function defined on
the set $\mathfrak{S}(\mathcal{H})$. All these results are based on
the following general observations. \vspace{5pt}

\begin{property}\label{ppm-t-1}
\textit{Let $f$ be a lower bounded lower semicontinuous function on
the set $\,\mathfrak{S}(\mathcal{H})$.}\medskip

A) \textit{The functions
$$
\check{f}(\rho)\doteq\inf_{\mu\in\mathcal{P}_{\{\rho\}}(\mathfrak{S}(\mathcal{H}))}\int
f(\sigma) \mu(d\sigma)\quad and\quad\hat{f}(\rho)\doteq
\sup_{\mu\in\mathcal{P}_{\{\rho\}}(\mathfrak{S}(\mathcal{H}))}\int
f(\sigma)\mu(d\sigma)
$$
are lower semicontinuous on the set $\,\mathfrak{S}(\mathcal{H})$.}\medskip

B) \textit{For an arbitrary state $\rho\in\mathfrak{S}(\mathcal{H})$
there exists a measure
$\mu^{f}_{\rho}\in\mathcal{P}_{\{\rho\}}(\mathfrak{S}(\mathcal{H}))$
at which the infimum in the definition of the value
$\check{f}(\rho)$ is achieved}. \textit{If $f$ is a concave function
then the measure $\mu^{f}_{\rho}$ can be chosen in
$\,\mathcal{P}_{\{\rho\}}(\mathfrak{S}_1(\mathcal{H}))$.}\medskip

C) \textit{The function $\check{f}$ coincides with the convex
closure $\,\overline{\mathrm{co}}f$ of the function $f$, while the
function $\hat{f}$ -- with the concave hull \footnote{the minimal
concave function majorizing the function $f$.} of the function $f$,
that is
$$ \hat{f}(\rho)=
\sup_{\{\pi_{i},\rho_{i}\}\in\mathcal{P}^{f}_{\{\rho\}}(\mathfrak{S}(\mathcal{H}))}\sum_{i}\pi_{i}f(\rho_{i}),
\quad\forall\rho\in\mathfrak{S}(\mathcal{H}).
$$}

\end{property}\vspace{5pt}

\textbf{Proof.} Assertion A, the first part of B and the second part
of C follow from Corollaries \ref{c-3} and \ref{c-5}.

If $f$ is a concave function  then optimality of the measure
$\mu^{f}_{\rho}$ implies optimality of any measure
$\nu\in\mathcal{P}_{\{\rho\}}(\mathfrak{S}_1(\mathcal{H}))$ such
that $\nu\succ\mu^{f}_{\rho}$ by Lemma \ref{order-ineq+} (existence
of such measure $\nu$ follows from Corollary \ref{ppm-c-n}).

Since the convex function $\check{f}$ is lower semicontinuous,  the
definition of the convex closure implies
$\check{f}(\rho)\leq\overline{\mathrm{co}}f(\rho)$ for all $\rho$ in
$\mathfrak{S}(\mathcal{H})$ . Since $\overline{\mathrm{co}}f$ is a
convex lower semicontinuous function majorized by $f$, Jensen's
inequality implies
$$
\overline{\mathrm{co}}f(\rho)\leq
\inf_{\mu\in\mathcal{P}_{\{\rho\}}(\mathfrak{S}(\mathcal{H}))}\int
\overline{\mathrm{co}}f(\sigma) \mu(d\sigma)\leq
\inf_{\mu\in\mathcal{P}_{\{\rho\}}(\mathfrak{S}(\mathcal{H}))}\int
f(\sigma) \mu(d\sigma)=\check{f}(\rho)
$$
for all $\rho$ in $\mathfrak{S}(\mathcal{H})$. It follows that
$\check{f}=\overline{\mathrm{co}}f$. $\square$\medskip

\textbf{Remark 1.} The analogies of assertion B and of the second
part of C in Proposition \ref{ppm-t-1} do not hold for the functions
$\hat{f}$ and $\check{f}$ correspondingly.

It is easy to construct a bounded lower semicontinuous function $f$
and a state $\rho_{0}$ such that the supremum in the definition of
the value $\hat{f}(\rho_{0})$ is not achieved. Indeed, let
$$
f(\rho)=\left\{
   \begin{array}{ll}
    \mathrm{Tr}\rho^{2}, & \rho\;\; \textup{is}\; \textup{a}\; \textup{mixed}\; \textup{state}\\
    0, & \rho\;\; \textup{is}\; \textup{a}\; \textup{pure}\; \textup{state}
    \end{array}\right.
$$
be a lower semicontinuous bounded function and
$\rho_{0}=\frac{1}{2\pi}\int_{0}^{2\pi}V_{t}|\psi\rangle\langle\psi|V_{t}^{*}dt$,
where $V_{t}$ is the unitary representation in the Hilbert space
$\mathcal{H}$ of the torus $\mathbb{T}$, identified with $[0,2\pi)$,
and $|\psi\rangle$ is an arbitrary vector in $\mathcal{H}$. Then
$\hat{f}(\rho_{0})=1$ and hence the supremum in the definition of
the value $\hat{f}(\rho_{0})$ is not achieved since $f(\rho)<1$ for
all $\rho$. To show that $\hat{f}(\rho_{0})=1$ it is sufficient to
note that the mixed state
$\,\rho_{\delta}=\delta^{-1}\int_{0}^{\delta}V_{t}|\psi\rangle\langle\psi|V_{t}^{*}dt\,$
tends to the pure state $|\psi\rangle\langle\psi|$ and hence
$f(\rho_{\delta})$ tends to $1$ as $\delta\rightarrow+0$.

To show that the infimum in the definition of the value
$\check{f}(\rho)$ can not be taken over the set
$\mathcal{P}^{\,f}_{\{\rho\}}(\mathfrak{S}(\mathcal{H}))$ one can
consider the von Neumann entropy
$H(\rho)=-\mathrm{Tr}\rho\log\rho\,$ (is easy to see that
$\,\overline{\mathrm{co}}H=\check{H}\equiv0\,$ while
$\,\mathrm{co}H(\rho)=+\infty\,$ for any state $\rho$ with
$H(\rho)=+\infty$, since the set of states with finite entropy is
convex). $\square$

If $f$ is a continuous bounded function on the set
$\mathfrak{S}(\mathcal{H})$ then Proposition \ref{ppm-t-1} can be
applied to the functions $f$ and $-f$ simultaneously resulting in
the following observation (which also directly follows from the
generalized Vesterstrom-O'Brien theorem described in Section 4).
\vspace{5pt}

\begin{corollary}\label{ppm-p-4}
\textit{Let $f$ be a continuous bounded function on the set
$\,\mathfrak{S}(\mathcal{H})$. The convex hull $\,\mathrm{co}f$ and
the convex closure $\,\overline{\mathrm{co}}f$ of the function $f$
coincide and the function $\,\overline{\mathrm{co}}f=\mathrm{co}f$
is continuous on the set $\,\mathfrak{S}(\mathcal{H})$.}
\end{corollary}\vspace{5pt}

Coincidence of $\,\overline{\mathrm{co}}f\,$ and $\,\mathrm{co}f\,$ holds
for any continuous function $f$ on a convex set $X$ if this set is
compact \cite[Corallary I.3.6]{Alf} or at least $\mu$-compact
\cite[Corallary 2]{P&Sh}, but it does not hold in general
\cite[Example 1]{P&Sh}. Continuity of the function
$\;\overline{\mathrm{co}}f=\mathrm{co}f\,$ is a corollary of stability
of the set $\,\mathfrak{S}(\mathcal{H})$, since even in
$\mathbb{R}^{3}$ there exist a convex compact set $X$ and a continuous
function $f$ on $X$ such that the function $\,\overline{\mathrm{co}}f=\mathrm{co}f$
is not continuous on $X$ \cite{Brien}.

As mentioned at the end of Remark 1 $\,\overline{\mathrm{co}}f\neq
\mathrm{co}f\,$ if the function $f$ is only lower semicontinuous and
lower bounded.

\subsection{On functions obtained via the generalized convex (concave) roof construction}

In the case  $\dim\mathcal{H}<+\infty$ the convex (concave) roof
extension to the set $\mathfrak{S}(\mathcal{H})$ of a function $f$
on the set
$\mathfrak{S}_{1}(\mathcal{H})=\mathrm{extr}\mathfrak{S}(\mathcal{H})$
of pure states is defined at a mixed state $\rho$ as the minimal
(maximal) value of $\sum_{i}\pi_{i}f(\rho_{i})$ over all
decompositions $\rho=\sum_{i}\pi_{i}\rho_{i}$ of this state into
finite convex combination of pure states \cite{U}. This extension is
widely used in quantum information theory, in particular, in
construction of entanglement monotones \cite{P&V}. The convex
(concave) roof extension has two natural generalizations to the case
$\dim\mathcal{H}=+\infty$ called in \cite{Sh-9} the
$\sigma$\nobreakdash-\hspace{0pt}convex (concave) roof and the
$\mu$\nobreakdash-\hspace{0pt}convex (concave) roof correspondingly
(the first extension is defined via all decompositions of a state
into countable convex combination of pure states while the second
one -- via all "continuous" decompositions corresponding to Borel
probability measures on the set of pure states with given
barycenter). In \cite{Sh-9} it is shown that it is the
$\mu$\nobreakdash-\hspace{0pt}convex roof that should be used for
construction of entanglement monotones in infinite dimensions.

For a given Borel function $f$ on the set
$\mathfrak{S}_{k}(\mathcal{H})$ of states of rank $\leq k$ one can
consider the generalized $\mu$\nobreakdash-\hspace{0pt}convex
(concave) roof defined at a state $\rho$ as the infimum (supremum)
of $\int f(\sigma) \mu(d\sigma)$ over all measures in
$\mathcal{P}_{\{\rho\}}(\mathfrak{S}_k(\mathcal{H}))$ \cite{Sh-11}.
The $\mu$-compactness and strong stability of the set
$\mathfrak{S}(\mathcal{H})$ (expressed in Theorems \ref{th-1} and
\ref{th-2}B respectively) imply the following observations
concerning properties of the generalized
$\mu$\nobreakdash-\hspace{0pt}convex (concave) roof.\medskip

\begin{property}\label{ppm-t-2}
\textit{Let $f$ be a lower bounded lower semicontinuous function on
the set $\,\mathfrak{S}_k(\mathcal{H})$,  $k\in\mathbb{N}$.}\medskip

A) \textit{The functions
$$
\check{f}_{k}(\rho)\doteq
\inf_{\mu\in\mathcal{P}_{\{\rho\}}(\mathfrak{S}_k(\mathcal{H}))}\int
f(\sigma) \mu(d\sigma)\quad and\quad \hat{f}_{k}(\rho)\doteq
\sup_{\mu\in\mathcal{P}_{\{\rho\}}(\mathfrak{S}_k(\mathcal{H}))}\int
f(\sigma) \mu(d\sigma)
$$
are lower semicontinuous  on the set $\,\mathfrak{S}(\mathcal{H})$.}\medskip

B) \textit{For an arbitrary state $\rho$ in
$\mathfrak{S}(\mathcal{H})$ there exists a measure
$\hat{\mu}^{f}_{\rho}$ in\break
$\mathcal{P}_{\{\rho\}}(\mathfrak{S}_k(\mathcal{H}))$ at which the
infimum in the definition of the value $\check{f}_{k}(\rho)$ is
achieved}.\medskip

C) \textit{The function $\hat{f}_{k}$ can be defined as follows
$$
\hat{f}_{k}(\rho)=
\sup_{\{\pi_{i},\rho_{i}\}\in\mathcal{P}^{\,a}_{\{\rho\}}(\mathfrak{S}_{k}(\mathcal{H}))}\sum_{i}\pi_{i}f(\rho_{i}),
\quad\forall\rho\in\mathfrak{S}(\mathcal{H}).
$$}
\end{property}

The all assertion of this proposition follow from Corollaries
\ref{c-3} and \ref{c-5}. \medskip

\textbf{Remark 2.} The analogies of the assertions B and C in
Proposition \ref{ppm-t-2} do not hold for the functions
$\hat{f}_{k}$ and $\check{f}_{k}$ correspondingly.

Below we construct a \emph{bounded} lower semicontinuous function
$f$ on the set of pure states and a state $\rho_{0}$ such that the
supremum in the definition of the value $\hat{f}_{1}(\rho_{0})$ is
not achieved.

Let $\mathcal{A}_{s}$ be the set of pure product states in
$\mathfrak{S}(\mathcal{H}\otimes\mathcal{H})$ and
$\rho_{0}\in\overline{\mathrm{co}}(\mathcal{A}_{s})$ be the
separable state in $\mathfrak{S}(\mathcal{H}\otimes\mathcal{H})$
constructed in \cite{H-Sh-W} such that any measure with the
barycenter $\rho_{0}$ has no atoms in the set $\mathcal{A}_{s}$. Let
$$
f(\rho)=\left\{
   \begin{array}{ll}
    \sup_{\sigma\in\mathcal{A}_{s}}\mathrm{Tr}\rho\sigma, & \rho\in
    \mathfrak{S}_1(\mathcal{H}\otimes\mathcal{H})\setminus\mathcal{A}_{s}\\
    0, & \rho\in\mathcal{A}_{s}.
    \end{array}\right.
$$
It is easy to see that the function $f$ is bounded and lower
semicontinuous on the set
$\mathfrak{S}_1(\mathcal{H}\otimes\mathcal{H})$. Since $f(\rho)<1$
for all $\rho$ in $\mathfrak{S}_1(\mathcal{H}\otimes\mathcal{H})$ to
show that the supremum in the definition of the value
$\hat{f}_{1}(\rho_{0})$ is not achieved it is sufficient to show
that $\hat{f}_{1}(\rho_{0})=1$. By Corollary \ref{c-1} there exists
a measure $\hat{\mu}_{0}$ (purely nonatomic) supported by the set
$\mathcal{A}_{s}$ such that $\textbf{b}(\hat{\mu}_{0})=\rho_{0}$. By
Corollary \ref{c-2} there exists a sequence $\{\hat{\mu}_{n}\}$ of
measures in
$\mathcal{P}^{\,a}_{\{\rho_{0}\}}(\mathfrak{S}_1(\mathcal{H}\otimes\mathcal{H}))$ weakly
converging to the measure $\hat{\mu}_{0}$. Since by Lemma
\ref{s-function} in the Appendix the bounded function
$g(\rho)=\sup_{\sigma\in\mathcal{A}_{s}}\mathrm{Tr}\rho\sigma$ is
continuous on the set $\mathfrak{S}(\mathcal{H}\otimes\mathcal{H})$,
the definition of weak convergence implies
\begin{equation}\label{limit-exp}
\lim_{n\rightarrow+\infty}\int g(\sigma)\hat{\mu}_{n}(d\sigma)=\int
g(\sigma)\hat{\mu}_{0}(d\sigma)=1.
\end{equation}
By the construction of the state $\rho_{0}$ for each $n$ all atoms
of the measure $\hat{\mu}_{n}$ lie in
$\mathfrak{S}_1(\mathcal{H}\otimes\mathcal{H})\setminus\mathcal{A}_{s}$
and hence
$$
\int g(\sigma)\hat{\mu}_{n}(d\sigma)=\int
f(\sigma)\hat{\mu}_{n}(d\sigma).
$$
This and (\ref{limit-exp}) imply
$\hat{f}_{1}(\rho_{0})=1$.\smallskip

To show that the infimum in the definition of the value
$\check{f}_{1}(\rho)$ can not be taken over the set
$\,\mathcal{P}^{\,a}_{\{\rho\}}(\mathfrak{S}_1(\mathcal{H}))\,$
consider the indicator function $\chi_{\bar{\mathcal{A}_{s}}}$ of
the set
$\bar{\mathcal{A}_{s}}=\mathfrak{S}_1(\mathcal{H}\otimes\mathcal{H})\setminus\mathcal{A}_{s}$,
where $\mathcal{A}_{s}$ is the set of pure product states in
$\mathfrak{S}(\mathcal{H}\otimes\mathcal{H})$ and
$\rho_{0}\in\overline{\mathrm{co}}(\mathcal{A}_{s})$ is the
separable state described in the above example. It is easy to see
that the bounded function $\chi_{\bar{\mathcal{A}_{s}}}$ is concave
and lower semicontinuous on the set
$\mathfrak{S}_1(\mathcal{H}\otimes\mathcal{H})$. We have
$$
\inf_{\mu\in\mathcal{P}_{\{\rho_{0}\}}(\mathfrak{S}_1(\mathcal{H}\otimes\mathcal{H}))}\int
\chi_{\bar{\mathcal{A}_{s}}}(\sigma)\mu(d\sigma)=0,
$$
while
$$
\inf_{\mu\in\mathcal{P}^{\,a}_{\{\rho_{0}\}}(\mathfrak{S}_1(\mathcal{H}\otimes\mathcal{H}))}\int
\chi_{\bar{\mathcal{A}_{s}}}(\sigma)\mu(d\sigma)=
\inf_{\{\pi_{i},\rho_{i}\}\in\mathcal{P}^{\,a}_{\{\rho_{0}\}}(\mathfrak{S}_1(\mathcal{H}\otimes\mathcal{H}))}
\sum_{i}\pi_{i}\chi_{\bar{\mathcal{A}_{s}}}(\rho_{i})=1,
$$
since by the construction of the state $\rho_{0}$ each countable
convex decomposition of this state does not contain states from
$\mathcal{A}_{s}$. $\square$ \medskip

If $f$ is a continuous bounded function on the set
$\mathfrak{S}_k(\mathcal{H})$ then Proposition \ref{ppm-t-2} can be
applied to the functions $f$ and $-f$ simultaneously resulting in
the following observation. \medskip

\begin{corollary}\label{ppm-p-5}
\textit{Let $f$ be a continuous bounded function on the set
$\,\mathfrak{S}_k(\mathcal{H})$, $k\in\mathbb{N}$.} \textit{Then the
functions $\check{f}_{k}$ and $\hat{f}_{k}$, introduced in
Proposition \ref{ppm-t-2}, are continuous on the set
$\,\mathfrak{S}(\mathcal{H})$  and
$$
\check{f}_{k}(\rho)=\inf_{\{\pi_{i},\rho_{i}\}\in\mathcal{P}^{\,a}_{\{\rho\}}(\mathfrak{S}_k(\mathcal{H}))}\sum_{i}\pi_{i}f(\rho_{i}),\quad
\hat{f}_{k}(\rho)=\sup_{\{\pi_{i},\rho_{i}\}\in\mathcal{P}^{\,a}_{\{\rho\}}(\mathfrak{S}_k(\mathcal{H}))}\sum_{i}\pi_{i}f(\rho_{i})
$$
for any state $\rho\in\mathfrak{S}(\mathcal{H})$.}
\end{corollary}

\subsubsection{The case $k=1$.}

Corollary \ref{ppm-p-5} with $k=1$ implies, in particular, the
following observation, which is nontrivial since the set
$\mathfrak{S}(\mathcal{H})$ is not compact (cf. Example 1 in \cite{P&Sh}).\medskip

\begin{corollary}\label{ppm-c-7}
\textit{An arbitrary continuous bounded function on the set
$\,\mathfrak{S}_1(\mathcal{H})$ can be extended to convex (concave)
continuous bounded function on the set $\mathfrak{S}(\mathcal{H})$.}
\end{corollary}\medskip

Corollary \ref{ppm-p-5} also implies the following criterion of
continuity of a convex closure of concave functions.\medskip

\begin{corollary}\label{ppm-c-9}
\textit{ Let $f$ be a concave lower bounded lower semicontinuous
function on the set $\,\mathfrak{S}(\mathcal{H})$. The convex
closure $\,\overline{\mathrm{co}}f$ of the function $f$ is bounded
and continuous on the set $\,\mathfrak{S}(\mathcal{H})$ if and only
if the function $f$ has bounded and continuous restriction to the
set
$\,\mathfrak{S}_1(\mathcal{H})=\mathrm{extr}\mathfrak{S}(\mathcal{H})$.
In this case
$$
\overline{\mathrm{co}}f(\rho)=\inf_{\{\pi_{i},\rho_{i}\}\in\mathcal{P}^{\,a}_{\{\rho\}}(\mathfrak{S}_1(\mathcal{H}))}\sum_{i}\pi_{i}f(\rho_{i}),
\quad\forall\rho\in\mathfrak{S}(\mathcal{H}).
$$
}
\end{corollary}\medskip

\textbf{Proof.} By the condition Proposition \ref{ppm-t-1}B implies
$\,\overline{\mathrm{co}}f=\check{f}_{1}$.

If the function $f$ is bounded and continuous on the set
$\mathfrak{S}_1(\mathcal{H})$ then by Corollary \ref{ppm-p-5} the
function $\,\overline{\mathrm{co}}f=\check{f}_{1}\,$ is bounded and
continuous on the set $\mathfrak{S}(\mathcal{H})$.

The converse assertion is trivial, since the functions $f$ and
$\,\overline{\mathrm{co}}f=\check{f}_{1}\,$ coincide on the set
$\mathfrak{S}_1(\mathcal{H})$. $\square$\medskip

Corollary \ref{ppm-p-5} gives a criterion of continuity of a convex
closure of the output entropy $H_{\Phi}=H\circ\Phi$ of a quantum
channel $\Phi$, which is an important characteristic of this channel
related, in particular, to its classical capacity.

Proposition \ref{ppm-t-2} and Corollary \ref{ppm-p-5} with $k=1$
play an  essential role in an infinite dimensional generalization of
the convex roof construction of entanglement monotones considered in
\cite{Sh-9}.

\subsubsection{The case $k\in N$.}

Corollary \ref{ppm-p-5} provides a special approximation technique
for concave lower semicontinuous lower bounded functions on the set
$\mathfrak{S}(\mathcal{H})$ proposed in \cite{Sh-11} and briefly
described below.

Let $f$ be a concave lower semicontinuous lower bounded function on
the set $\mathfrak{S}(\mathcal{H})$ having continuous restrictions
to the set $\mathfrak{S}_k(\mathcal{H})$ for each $k\in N$. As a
simple nontrivial example of such function one can consider the
Renyi entropy of order $p\in(0,1]$, in particular, the von Neumann
entropy.

By Corollary \ref{ppm-p-5} the sequence $\{\hat{f}_{k}\}$ consists
of continuous concave bounded functions on the set
$\mathfrak{S}(\mathcal{H})$ such that
\begin{equation*}
\hat{f}_{k}\leq f\quad
\textup{and}\quad\hat{f}_{k}|_{\mathfrak{S}_k(\mathcal{H})}=f|_{\mathfrak{S}_k(\mathcal{H})}
\end{equation*}
(these relations are obtained by means of Jensen's inequality).

By  lower semicontinuity of the function $f$ the increasing sequence
$\{\hat{f}_{k}\}$ pointwise converges to the function $f$, t.i.
$f=\sup_k\hat{f}_{k}$ \cite[Section 4]{Sh-11}.\medskip

For example, if $\,f=H\,$ is the von Neumann entropy, then
$\{\hat{H}_{k}\}$ is an increasing sequence of continuous concave
unitary invariant functions such that
$$
\hat{H}_{k}|_{\mathfrak{S}_k(\mathcal{H})}=H|_{\mathfrak{S}_k(\mathcal{H})}\quad
\textup{and}\quad
\sup_k\hat{H}_{k}(\rho)=H(\rho),\;\;\rho\in\mathfrak{S}(\mathcal{H}).
$$

The sequence $\{\hat{H}_{k}\}$ can be used to obtain necessary and
sufficient conditions for local continuity of the von Neumann
entropy by exploiting well known relations between uniform
convergence of an increasing sequence of continuous functions and
continuity of the limit function \cite{Sh-11}. Applicability of
these conditions is based on possibility to express the difference
$H-\hat{H}_{k}$ via the relative entropy:
$$
\Delta_k(\rho)=H(\rho)-\hat{H}_{k}(\rho)=\inf_{\{\pi_{i},\rho_{i}\}\in\mathcal{P}^{\,a}_{\{\rho\}}(\mathfrak{S}_{k}(\mathcal{H}))}
\sum_{i}\pi_{i}H(\rho_{i}\|\rho).
$$
Properties of the function $\rho\mapsto\Delta_k(\rho)$ is studied in
detail in \cite[Lemma 8]{Sh-11}.\medskip

The sequence $\{\hat{H}_{k}\}$ can be also used to construct an
increasing sequence of continuous entanglement monotones providing
approximation of the Entanglement of Formation \cite[Section
6.3]{Sh-9}.

\section{Appendix}

\begin{property}\label{indicator}
\textit{Let  $\mathcal{A}$ be an arbitrary closed subset of
$\,\mathfrak{S}_1(\mathcal{H})$. The indicator function of the set
$\mathcal{A}$ coincides with the pointwise limit of the decreasing
sequence of continuous convex functions
$\,f_{n}(\rho)=1-\sqrt[n]{1-g_{\mathcal{A}}(\rho)}\,$, where}
$$
g_{\mathcal{A}}(\rho)=\sup_{\sigma\in\mathcal{A}}\mathrm{Tr}\rho\sigma,\quad\rho\in\mathfrak{S}(\mathcal{H}).
$$
\end{property}\medskip
This proposition follows from Lemma \ref{s-function} below and
concavity of the increasing function $\sqrt[n]{x}$.\medskip

\begin{lemma}\label{s-function} \textit{The function $g_{\mathcal{A}}$ is
convex and continuous on the set $\,\mathfrak{S}(\mathcal{H})$.}
\end{lemma}\medskip

\textbf{Proof.} Convexity and lower semicontinuity of the function
$g_{\mathcal{A}}$ follows from its representation as the least upper
bound of the family
$\{\mathrm{Tr}\rho\sigma\}_{\sigma\in\mathcal{A}}$ of bounded
continuous affine functions on $\mathfrak{S}(\mathcal{H})$.

Suppose that the function $g_{\mathcal{A}}$ is not upper
semicontinuous. This implies existence of a sequence $\{\rho_{n}\}$
of states converging to some state $\rho_{0}$ such that
\begin{equation}\label{b-l-s}
\lim_{n\rightarrow+\infty}g_{\mathcal{A}}(\rho_{n})>g_{\mathcal{A}}(\rho_{0}).
\end{equation}
Let $\mathfrak{A}=\{|\varphi\rangle\in\mathcal{H}\,|\,
|\varphi\rangle\langle\varphi|\in \mathcal{A}\}$ be subset of
$\mathcal{H}$ and $\overline{\mathfrak{A}}$ be its  closure in the
weak topology in $\mathcal{H}$. Lemma 2 on p.284 in \cite{K&F}
\footnote{Let $\{x_{n}\}$ be an arbitrary sequence of vectors in a
Hilbert space weakly converging to the vector $x$ and $A$ be an
arbitrary compact operator. Then $\lim_{n\rightarrow+\infty}\langle
x_{n}|A|x_{n}\rangle=\langle x|A|x\rangle$.} implies
\begin{equation}\label{sup-exp}
g_{\mathcal{A}}(\rho_{0})=\sup_{\sigma\in\mathcal{A}}\mathrm{Tr}\rho_{0}\sigma=
\sup_{\varphi\in\mathfrak{A}}\langle\varphi|\rho_{0}|\varphi\rangle=
\sup_{\varphi\in\overline{\mathfrak{A}}}\langle\varphi|\rho_{0}|\varphi\rangle.
\end{equation}

For arbitrary $\varepsilon>0$ and arbitrary $n$ there exists a
vector $\varphi^{\varepsilon}_{n}\in\mathfrak{A}$ such that
$\langle\varphi^{\varepsilon}_{n}|\rho_{n}|\varphi^{\varepsilon}_{n}\rangle>
g_{\mathcal{A}}(\rho_{n})-\varepsilon$. Since the unit ball of the
space $\mathcal{H}$ is compact in the weak topology we can find a
subsequence $\{\varphi^{\varepsilon}_{n_{k}}\}_{k}$ of the sequence
$\{\varphi^{\varepsilon}_{n}\}_{n}$ weakly converging to some vector
$\varphi^{\varepsilon}\in\overline{\mathfrak{A}}$.  By Lemma 2 on
p.284 in \cite{K&F} the sequence
$\{\langle\varphi^{\varepsilon}_{n_{k}}|\rho_{0}|\varphi^{\varepsilon}_{n_{k}}\rangle\}_{k}$
converges to
$\langle\varphi^{\varepsilon}|\rho_{0}|\varphi^{\varepsilon}\rangle$
as $k$ tends to the infinity. This and the estimation
$|\langle\varphi^{\varepsilon}_{n_{k}}|\rho_{n_{k}}-\rho_{0}
|\varphi^{\varepsilon}_{n_{k}}\rangle|\leq
\|\rho_{n_{k}}-\rho_{0}\|_{1}$ imply
$$
\lim_{k\rightarrow+\infty}g_{\mathcal{A}}(\rho_{n_{k}})\leq
\lim_{k\rightarrow+\infty}\langle\varphi^{\varepsilon}_{n_{k}}|\rho_{n_{k}}
|\varphi^{\varepsilon}_{n_{k}}\rangle-\varepsilon=
\langle\varphi^{\varepsilon}|\rho_{0}|\varphi^{\varepsilon}\rangle-\varepsilon\leq
g_{\mathcal{A}}(\rho_{0})-\varepsilon,
$$
where the last inequality follows from (\ref{sup-exp}). But this
contradicts to assumption (\ref{b-l-s}) since $\varepsilon$ is
arbitrary. $\square$\bigskip

\textbf{Acknowledgments.} The author is grateful to A. S. Holevo for
the help and useful discussion.  This work is partially supported by
the program "Mathematical control theory" of Russian Academy of
Sciences, by the federal target program "Scientific and pedagogical
staff of innovative Russia" (program 1.2.1, contract P 938), by the
analytical departmental target program "Development of scientific
potential of the higher school 2009-2010" (project 2.1.1/500) and by
RFBR grant 09-01-00424-a.

\end{document}